\documentclass{article}
\begin{document}

\begin{center}
{\Large  NEW CASIMIR ENERGY CALCULATIONS }
\end{center}
\begin{center}
H. Ahmedov
\\
Feza Gursey Institute, 81220, Istanbul, Turkey.
\\
E-mail: hagi@gursey.gov.tr
\end{center}
\begin{center}
I. H. Duru
\\
Izmir  Institute of Technology, 35430,  Izmir, Turkey.
\\
E-mail: ihduru@galois.iyte.edu.tr
\end{center}
\vspace{1cm} \noindent
 New Casimir energy results for massless
scalar field in some 3 -dimensional cavities are presented. We
attempted to discuss the correlation between the sign and the
magnitude of the energy and the shape of the cavities.

\vspace{1cm} \noindent  {\bf 1. Introduction}
\\
Sign of the Casimir energy known to be dependent on the dimension,
topology and the shape of the geometry. In this note we present some
new exact results for massless scalar fields in three dimensional
cavities with the trivial topology. We then compare the known
Casimir energy values for several three dimensional cavities. The
conclusion we arrived is that  the existence of the corners lowers
the vacuum energy.

\vspace{1cm}\noindent {\bf 2. New Casimir Energy Results in  Some
3-dimensional Cavities}
\\
In this section we present Casimir Energies for massless scalar
field in some 3-dimensional cavities. These cavities are rather
special regions, for all of them are fundamental domains for some
crystallographic group generated by reflections with respect to the
boundary walls. This property enables us to obtain the wave
functions satisfying the Dirichlet boundary conditions and then the
correct energy spectrum.
\\
(i) \underline{A Pyramidal Cavity}
\\
The region is defined by the planes
\begin{equation}
P_1: z=x, \ P_2: \ y=0, \ \ P_3: y=z, \ \  P_4: z=a.
\end{equation}
This is the fundamental domain of the group of order 48 generated by
the reflection with respect to the above planes \cite{1}. The
Casimir energy for massless scalar field in this cavity is ( in
$\hbar=c=1$ units )
\begin{equation}
    E_{pyr}\simeq \frac{0.069}{a}> 0
\end{equation}
(ii) \underline{A Conical Cavity}
\\
The conical cavity we consider is the one with   height $h=a$ and
with a very special  opening angle $\beta = \arcsin\frac{1}{3}$
\cite{2}. The crystallographic group which admits this cavity as the
fundamental domain is the Tetrahedral group. The Casimir  energy due
to the fluctuation of the massless scalar field is \cite{2}
\begin{equation}
    E_{con}\simeq \frac{0.080}{a}> 0
\end{equation}
(iii) \underline{ Triangular Cylinders}
\\
Three kind of triangle are the fundamental domain of some
crystallographic groups in the plane. These are equilateral,
right-angled isosceles and the right-angled triangle which is the
half of the equilateral one \cite{3}. Here we give the results for a
cylindrical cavity of  height $b$ and with equilateral triangular
cross-section of edges a. Three possibilities are distinguished:
\\
a) For $b> a$
\begin{equation}
    E_{tri}\simeq -\frac{0.053}{a}+ \frac{(0.029)b}{a^2}.
\end{equation}
\\
b) For  $a> b>\frac{a}{\sqrt{2}}\simeq 0,7 a$
\begin{equation}
    E_{tri}\simeq \frac{1}{2}(-\frac{0.013}{b}+\frac{(0.011)a}{b^2} +\frac{0.093}{a}-
    \frac{(0.048)b}{a^2}.
\end{equation}
\\
c) For $b< \frac{a}{\sqrt{2}}$
\begin{equation}
    E_{tri}\simeq -\frac{0.039}{b}+ \frac{(0.014)a}{b^2}.
\end{equation}
The energy for height $a$ is
\begin{equation}
    E_{tri}\simeq \frac{0,022}{a}.
\end{equation}
\\
\vspace{1cm} {\bf 3. Some Known Casimir Energy Results for
3-dimensional Cavities }
\\
(i) Casimir energy for the spherical cavity of radius a is \cite{4}
\begin{equation}
    E_{ball}\simeq \frac{0,045}{a}.
\end{equation}
\\
(ii) Coming to the cylinders with rectangular cross-sections we list
the results for the ones of square cross-section of edges a and of
height b \cite{5}:
\begin{equation}
    E_{rect}\simeq -\frac{0.013}{a}+ \frac{(0.011)b}{a^2} \ \ \  for
    \ b>a
\end{equation}
and
\begin{equation}
    E_{rect}\simeq -\frac{0.013}{b}+ \frac{(0.011)a}{b^2} \ \ \  for
    \ b<a
\end{equation}
For the cube of edges a we have
\begin{equation}
    E_{cub}\simeq -\frac{0.002}{a}<0
\end{equation}
which is very small.

\vspace{1cm}\noindent {\bf 4. Discussion}
\\
To have a meaningful comparison of the results we consider the
cavities of equal volumes. Such an approach may help us to
understand the shape dependence of the energy. The positive energies
for the pyramidal, conical  and triangular cylinder cavities can be
expressed in terms of energy of spherical cavity of the same volume
as:
\begin{eqnarray}
  E_{pyr} &\simeq& 0,51 E_{sph} \\
  E_{con} &\simeq& 0,54 E_{sph} \\
 E_{tr} &\simeq& 0,11 E_{sph}
\end{eqnarray}
The energy for the cube ( which is negative but very close to zero
) of the same volume is
\begin{equation}
    E_{cub}\simeq -0,0003 E_{sph}
\end{equation}
It seems that corners of the cavity reduces the energy. If we think
in terms of path integral picture we can say that the paths hitting
the corners cannot bounce back, but disappear. Thus we can think
that corners in the cavities reduces the phase space volume, and
then reduces the vacuum energy.

\vspace{1cm} \noindent {\bf Acknowledgments} The authors  H. Ahmedov
and I. H. Duru thanks the Turkish Academy of Sciences (TUBA) for its
support.


\begin{thebibliography}{99}
\bibitem{1} H. Ahmedov and I.H. Duru, {\it J. Math. Phys.} {\bf 46 },022303
(2005).
\bibitem{2} H. Ahmedov and I.H. Duru, {\it J. Math. Phys.} {\bf 46 },022304 (2005).
\bibitem{3} H. Ahmedov and I.H. Duru, in preparation. See also H. Ahmedov and I.H. Duru, {\it J. Math. Phys.} {\bf 45
}, 965 (2004 ).
\bibitem{4} T. H. Boyer, {\it Phys. Rev.} {\bf 174}, 1764 (1968);
B. Davies, {\it J.  Math. Phys.} , {\bf 13}, 1324 (1972); R. Balian
and B. Duplantier, {\it Ann. Phys.}, {\bf 104}, 300 (1978); J
Schwinger, L.L. De Raad and K. A. Milton, {\it Ann. Phys.}, {\bf
115}, 1 (1978); {\bf 115}, 388 (1978).
\bibitem{5} See for example V.M. Mostapanenko and N.N. Trunov "The
Casimir Effect and its Applications" Oxford Univ. Press, New York
(1997 ) and references therein.
\end{thebibliography}
\end{document}